\documentclass[aps,twocolumn]{revtex4}
\usepackage{amsmath,amssymb}
\input epsf
\usepackage{verbatim}
\usepackage{graphicx}
\usepackage{subfigure}
\usepackage{epstopdf}
\usepackage{bm}
\usepackage{xcolor}
\def\pmb#1{\setbox0=\hbox{#1}
\kern-.025em\copy0\kern-\wd0 \kern-.05em\copy0\kern-\wd0
\kern-.025em\raise.0433em\box0}

\def\ri  {{\rm i}}

\newcommand{\beq}{\begin{equation}}
\newcommand{\eeq}{\end{equation}}
\newcommand{\ba}{\begin{eqnarray}}
\newcommand{\ea}{\end{eqnarray}}

\input epsf

\usepackage[english]{babel}
\begin{document}

%\doublespace

\title[]{Robust collimated beaming in 3D acoustic sonic crystals}
%%{Acoustic endoscopic effect
%arising from the minima of an optical band
%%via dynamic anisotropy in a cubic array of rigid inclusions}
\author{A.~L. Vanel$^{1}$, M. Dubois$^{2,3}$, C. Tronche$^{4}$, S. Fu$^{5}$, Y.-T. Wang$^{6,7}$, G. Dupont$^{8}$, A.~D. Raki\'c$^{9}$, K. Bertling$^{9}$,
%Y. Achaoui$^{7}$, T. Antonakakis$^{8}$,
R. Abdeddaim$^{2}$, S. Enoch$^{2}$, R.~V. Craster$^{6,10}$, G. Li$^{11}$, S. Guenneau$^{10}$, J. Perchoux$^{4}$}
\affiliation{$^1$ CERN, 1211 Geneva 23, Switzerland}
\affiliation{$^2$ Aix$-$Marseille Univ, CNRS, Centrale Marseille, Institut Fresnel, Marseille, France}
\affiliation{$^3$ Multiwave Imaging SAS, Marseille, France}
\affiliation{$^4$ LAAS-CNRS, Universit\'e de Toulouse, CNRS, INP, Toulouse, France}
\affiliation{$^5$ Eastern Institute for Advanced Study, NingBo, China}
\affiliation{$^6$ Department of Mathematics, Imperial College London, London SW7 2AZ, UK}
\affiliation{$^{7}$ Photonics Initiative, Advanced Science Research Center, City University of New York, NY 10031, USA}
\affiliation{$^8$ Aix$-$Marseille Univ, CNRS, Centrale Marseille, IRPHE, Marseille, France}
\affiliation{$^9$ School of Information Technology
and Electrical Engineering, The University of Queensland, Brisbane, 4072, Australia}
\affiliation{$^{10}$ UMI 2004 Abraham de Moivre-CNRS, Imperial College London, London SW7 2AZ, UK}
%\affiliation{$^7$ Universit\'e de Franche-Comt\'e, CNRS, ENSMM, FEMTO-ST, 25000 Besancon, France} 
%\affiliation{$^8$ Multiwave Technologies AG, 3 Chemin du Pre Fleuri, 1228 Geneva, Switzerland}
\affiliation{$^{11}$ Department of Mathematics, The University of Hong Kong, Pokfulam Road, Hong Kong.}

\begin{abstract}

We demonstrate strongly collimated beaming, at audible frequencies, in a three-dimensional acoustic phononic crystal where the wavelength is commensurate with the crystal elements; the crystal is a seemingly simple rectangular cuboid constructed from closely-spaced spheres, and yet demonstrates rich wave phenomena  acting as a canonical three-dimensional metamaterial. 
 We employ theory, numerical simulation and experiments to design and interpret this collimated beaming phenomenon and use a crystal consisting of a finite rectangular cuboid array of $4\times 10\times 10$ polymer spheres $1.38$~cm in diameter in air, arranged in a primitive cubic cell with the centre-to-centre spacing of the spheres, i.e. the  pitch, as  $1.5$~cm.
%and experimentally characterized
%following some asymptotic and numerical models used as a guidance for the design of an acoustic endoscope operating at $14.2$ kHz and $18$ kHz.
Collimation effects are observed in the time domain for chirps with central frequencies at $14.2$~kHz and $18$~kHz, and we deployed a laser feedback interferometer or Self-Mixing Interferometer (SMI) -- a recently proposed technique to observe complex acoustic fields -- that enables experimental visualisation of the pressure field both within the crystal and outside of the crystal. Numerical exploration using a higher-order multi-scale finite element method designed 
%~\cite{hmsfem_acoustic,hmsfem_elastic}
for the rapid and detailed simulation of 3D wave physics further confirms these collimation effects and cross-validates with the experiments.
%for
%either a point source or
%a time-harmonic source with compact support placed in the near field of the sonic crystal at $14.2$ kHz and $18$ kHz. 
Interpretation follows using High Frequency Homogenization and Bloch analysis whereby the different origin of the collimation at these two frequencies is revealed by markedly different isofrequency surfaces of the sonic crystal.
%At $14.2$ kHz, high-frequency homogenization confirms
%that the endoscopic effect can be attributed to a 
%an unexpected flat ellipsoidal surface leads to the required large effective anisotropy.
%at $14.2$ kHz and $18$ kHz.
%captured by a high-frequency homogenization model 
%captures the main features of the second band of a Bloch diagram
%related to a quadratic behaviour near the high-symmetry point X, forcing ray trajectories along the $(1 0 0)$ crystallographic direction. 
%is preserved for cubic inclusions with same volumes as spheres.
%We stress that this is not a self-collimation effect, but rather waves bouncing back and forth along ray trajectories akin to negative refraction in the sonic crystal.


\end{abstract}

%\label{firstpage}
\maketitle

\section{Introduction}

A challenge for acoustic phononic crystals is to generate exotic wave phenomena at low-frequencies, i.e. at audible frequencies, and thereby obtain precise control over sound in the human hearing range. Motivated by arrays of closely-spaced two-dimensional crystals \cite{vanel2017asymptotic}, that have the particularly appealing feature that they operate as subwavelength tuneable metamaterials and allow the cut-off frequency for the acoustic branch to be lowered thereby pulling the entire acoustic branch subwavelength, 
 we explore the potential of closely-packed three dimensional spheres. This closely-packed limit is of interest as, in two-dimensions the analogous cylinder lattices create a network of coupled Helmholtz-like resonators with large voids coupled to each other by narrow gaps and in this way a precise asymptotic approach generates a discrete mass-spring analogy without any need for lumped parameters. 
 A natural extension is to three-dimensions and to see whether closely-packed arrangements of spheres have similar metamaterial behaviours - the situation is considerably more complex as numerical simulations of meaningfully large finite arrays are hampered by the need to resolve the thin air-filled regions between the spheres, the experiments are also difficult as it is hard to image within a crystal, and the asymptotic matching methods~\cite{vanel2017asymptotic} employed no longer are valid - indeed there are regions where the spheres nearly touch, but equally other regions where the acoustic medium is not as constrained as in two-dimensions - and so there is not a direct analogy with the two-dimensional system of cylinders. 

In this article we take a crystal that has close packing, and use a specially designed numerical scheme to overcome the computational issues \cite{hmsfem_acoustic}, an experimental optical interferometry approach \cite{bertling_OE_2014}, developed by some of the team, to enable the imaging, and develop the physical modelling and intuition of closely spaced media; bringing all this together then gives a coherent design capability that we use to create collimated beaming and we then demonstrate that closely-packing a simple phononic crystal can enable it to have metamaterial behaviours in the sense that each element within the crystal is subwavelength and yet critically affects the macroscale behaviour.
 There is a vast literature on acoustic metamaterials, \cite{craster12book,kadic2013metamaterials,cummer2016controlling}, including analysis of their 3D effective properties \cite{dupont2019analysis,liu06a,yves2021extreme} and the fabrication of three dimensional mechanical \cite{buckmann2012tailored} and acoustic \cite{liu2020three} metamaterials with complex geometry is done routinely nowadays. There is, however, a  richness of wave phenomena in simple three-dimensional crystals that has not yet been fully explored, and exploited; an example being the acoustic sonic crystal designed to achieve a  flat lens effect \cite{dubois2019acoustic}. 

We draw upon the extensive literature, primarily in two dimensions and in electromagnetism, using iso-frequency contours to predict extreme dynamic anisotropy; polygonal contours leading to self-collimation \cite{yu03a,chigrin2003self,prather07a}. Self-collimation is a powerful concept,
 readily observed in 3D for discrete  mass-spring systems  \cite{vanel2016asymptotics}, 
but apart from some notable examples in electromagnetism, i.e. \cite{lu2006experimental}, its use in three dimensional continuum models has been limited. \cite{Prather2005SelfcollimationI3,lu2006experimental} identify degenerate polarised modes in electromagnetism that have distinctive isofrequency surfaces allowing for strong internal beam formation within a crystal and here we identify the existence of acoustic modes, with similar surfaces, that we employ to create collimated beams exiting a phononic crystal. 
 The richness of behaviours possible through dynamic anisotropy is also illustrated in 
 \cite{belov2005homogenization} for electromagnetic crystals notably cubic arrays of split ring resonators and also \cite{gralak2000anomalous,luo2002all,ceresoli2015dynamic} for dynamic anisotropy in 2D dielectric photonic crystals. We take these approaches into acoustics and demonstrate that, for sources exterior to a crystal, we can use a lattice medium to generate wide or narrow acoustic beams. By doing so we highlight the versatility and tunability that is achievable without introducing complexity to a basic phononic structure and bring numerical, experimental and theoretical analysis to bear to provide a comprehensive analysis of the collimation effects created.

The plan of the article is as follows: 
The experimental method implemented in this work is based upon the acousto-optic effect that links the refractive index of the medium to the local pressure variations induced by the acoustic waves that propagate~\cite{ciddor1996}. A dedicated imaging system has been designed and built that is based on the Self-Mixing Interferometry (SMI) sensing methodology~\cite{taimre2015,bertling_OE_2014} providing a lightweight, non-intrusive and time-resolved measurement tool; our experimental set-up and results are described in section \ref{sec:exp}. 

Our array of $400$ spherical rigid spheres of diameter $s=1.38$~cm for airborne acoustic waves, with a center-to-center spacing $a=1.5$~cm, are not extremely tightly packed but simulations are none the less challenging in terms of resolving the computations in the narrowest gaps, memory storage constraints, whilst additionally facing the challenge of simulating the fields accurately, and fast, in three-dimensions. 
%; there is also numerical complexity in representing the dispersion surfaces in three-dimensions which is considerably more demanding that for two-dimensional doubly periodic structures. 
 Given the challenges this poses for standard finite element methods we opt to use higher-order multi-scale Finite Element Method (HMsFEM~\cite{hmsfem_acoustic,hmsfem_elastic}) that are designed to accurately solve the time-harmonic wave equation with a relatively coarse grid thanks to well-chosen multi-scale basis functions. This makes it possible to rapidly explore wave phenomena for different sonic crystal configurations and for sources at different frequencies: the endoscope phenomenon we illustrate in this article was achieved after fine tuning the geometric and acoustic parameters of a 3D crystal. 
Section \ref{sec:HMsFEM} is devoted to the higher-order Multi-scale Finite Element Method used to numerically solve the scattering problem %\eqref{eq:helmholtz}-\eqref{eq:nuemann} 
 posed in the whole space, when a source term is present
 %in the right-hand side of (\ref{eq:helmholtz})
 and outgoing wave conditions are enforced with Perfectly Matched Layers (PML). This approach allows us to perform highly accurate numerical simulations rapidly in three-dimensions with a coarse mesh. 
%We briefly review the numerical approach in section \ref{sec:HMsFEM}. 

In section \ref{sec:HFH} an effective model that works beyond the quasi-static limit is employed to reveal  extreme dynamic anisotropy of the crystal at specific frequencies, and this explains the origin of the collimation effect observed at $14$ kHz. The experimental characterization of the crystal in the time domain requires a specially designed setup to observe not only the dynamics of the pressure field outside of the crystal, but more importantly, inside of it. For the modelling we use the (non-dimensionalised) Helmholtz equation that holds for time-harmonic acoustic waves: 
\beq
 \frac{\partial^2 u}{\partial x_1^2}+\frac{\partial^2
   u}{\partial x_2^2}+ \frac{\partial^2 u}{\partial x_3^2}+\Omega^2 u=0,
\label{eq:helmholtz}
\eeq
 for $u({\bf x})=u(x_1,x_2,x_3)$ with $r=\sqrt{x_1^2+x_2^2+x_3^2}>s$ (i.e. everywhere outside of the spheres in the absence of a source) with $\Omega=\omega {a}/c$, with $a$ the cubic array pitch (in unit of meters) and
 where $\omega$ is the angular frequency (in unit of rad/s) and $c=340$~m/s. Consistent with the acoustically sound-hard boundary conditions on each sphere, we take Neumann boundary conditions, the normal derivative of the function being zero, to hold on the surface of each sphere
\beq
\frac{\partial u}{\partial \bf{n}}\mid_{r=s}=0,
\label{eq:nuemann}
\eeq
where ${\bf n}$ denotes the outwards pointing vector to the spheres. Section \ref{sec:Blochtheory} also presents the Bloch eigenvalue problem \eqref{eq:helmholtz}-\eqref{eq:nuemann} posed in a periodic cell with Floquet-Bloch conditions. Here, we use a standard finite element scheme  implemented in the Comsol Multiphysics package, since the computational domain is just a cubic cell containing a single  sphere.

%Indeed, our aim here is to provide evidence of the origin of an endoscopic effect observed at two specific frequencies of $14.2$ kHz and $18$ kHz.  

%However, we also make use of the High-Frequency Homogenization (HFH) method to finely approximate dispersion surfaces as computational resources required to plot dispersion surfaces in triply periodic structures might be too demanding for research groups without supercomputers at hand. Our aim is to describe a computational method, coined unfolding IBZ method, which enables to only compute dispersion curves, without missing important features otherwise encompassed in (4D) dispersion surfaces.

%Besides, our unfolding IBZ method can be applied with some {\it ad hoc} changes to other areas of physics whenever some periodicity occurs,
%such as solid-state physics~\cite{kittel96a}, water waves, elastodynamic waves or in the theory of composites~\cite{milton02a,movchan02a}.

Plotting the dispersion relations around the edges of the Brillouin zone~\cite{brillouin53a,kittel96a} is broadly sufficient to identify the extent of stop-band as the maxima and minima almost
always occur there \cite{joannopoulos08a}: some mathematical literature 
\cite{harrison07a} explicitly constructs counter-examples,
nonetheless it is almost always the case that the edges of the
Brillouin zone contain the essential information sufficient for most
purposes. However, as we described in our previous work \cite{dubois2019acoustic} one can actually
overlook a mode, or modes, that arise from a path within the Brillouin
zone and that this missing mode can be of interest: in \cite{dubois2019acoustic} it was shown to be responsible for the  lensing effect they found. We therefore analyse the Fermi surfaces to get a fuller picture of the wave phenomena responsible for the observed self-collimation effects at $14.2$ and $18$~kHz.

Finally some concluding remarks are drawn together in section \ref{sec:conclude}.

\section{Experiments}
\label{sec:exp}

\begin{figure}
%\vspace{2.0cm}
\includegraphics[width=8cm]{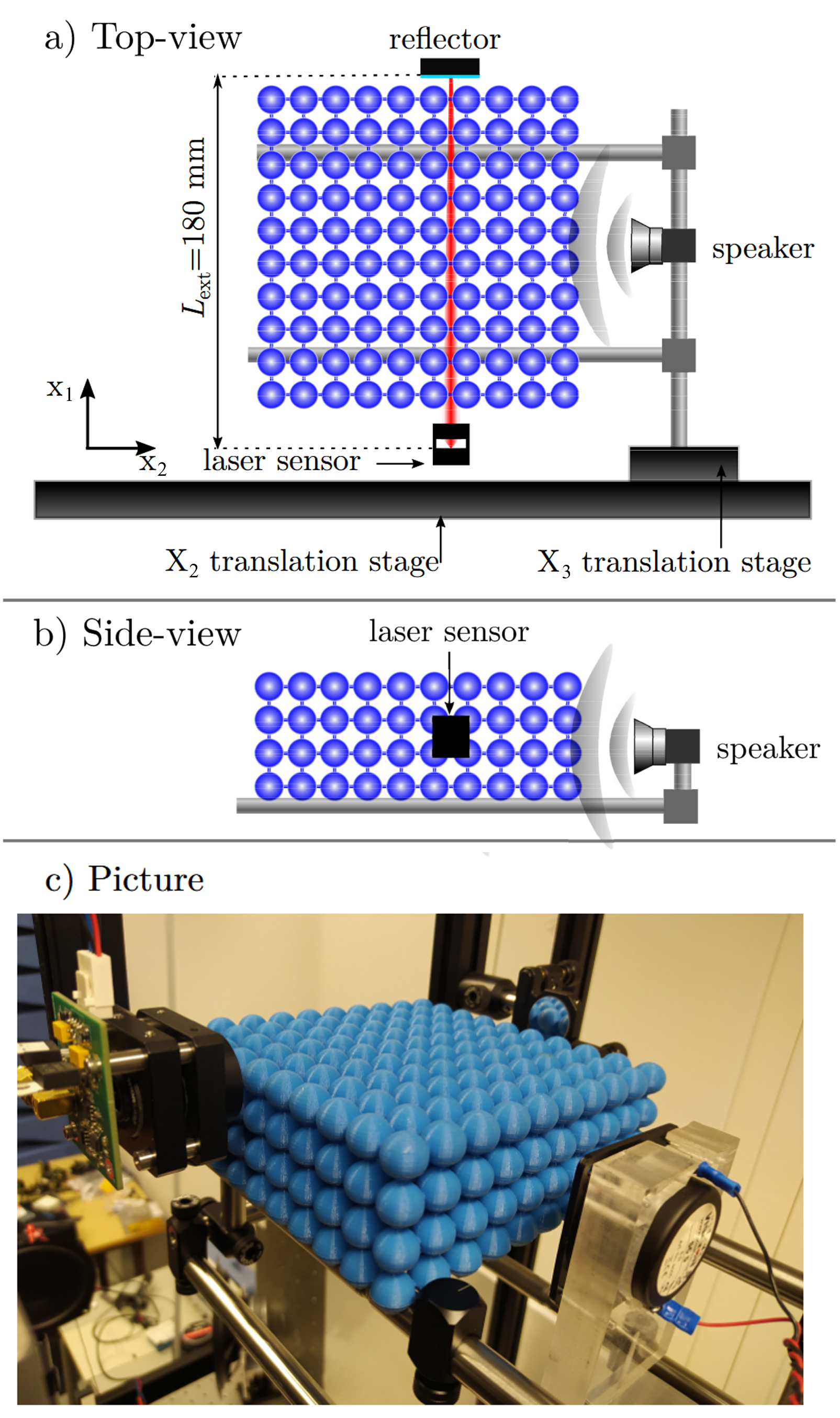}
\caption{Experimental setup: The crystal and the associated acoustic source are mounted on a rigid structure built with metallic rods. The structure is moved along $X_2$ and $X_3$ axis using motorized translation stages while the sensor and the distant reflector remain in a fixed position. a) Top-view. b) Side view. c) Photograph of the setup. }
\label{figaplsetup}
\end{figure}

%\subsection{Experimental setup}
The finite crystal used here is a rectangular cuboid array of 400 (4$\times$10$\times$10) 3D-printed polymer spheres 1.38~cm in diameter with a centre-to-centre spacing $a=1.5$~cm.
A commercial loudspeaker (Visaton SC5) is set 16~mm from the crystal -- which corresponds roughly to the wavelength of the acoustic wave -- and facing the thinner facet of the crystal (see~Fig.~\ref{figaplsetup}). It is driven with square pulses of sinusoidal signals with frequencies 14~kHz and 18~kHz. The pulses have a duration of 12~periods of the sinus and a spacing in time of 21~ms thus avoiding echoes appearing in the measurements while producing a relatively large frequency band signal. 

\begin{figure}[h!]
\includegraphics[width=7.5cm]{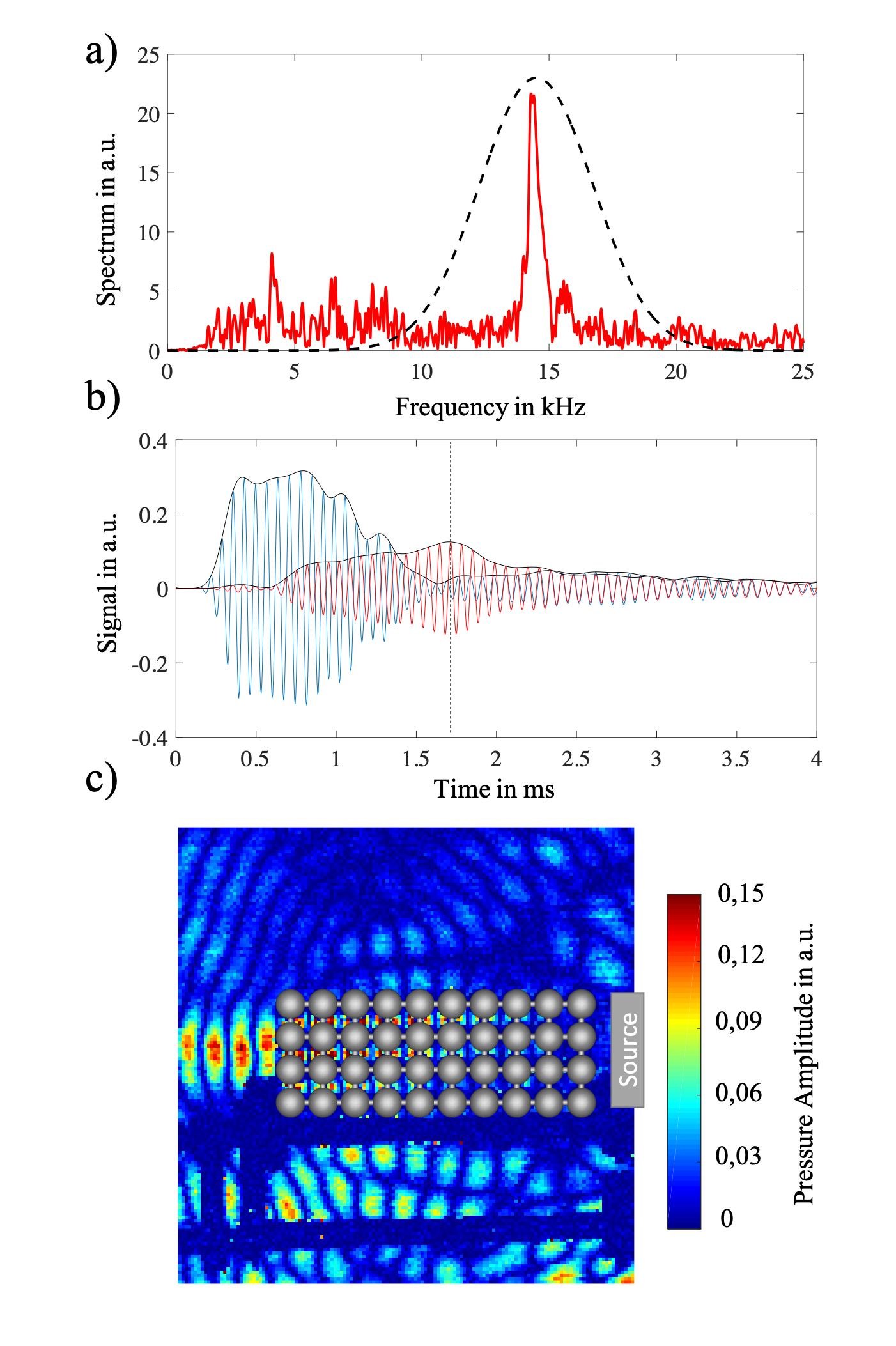}
\caption{Experimental observation at 14~kHz showing endoscopic effect: (a) Spectrum of the transmitted signal for an excitation centered at 14~kHz; the dotted line denotes the Gaussian window used. (b) Transmitted time trace in solid red with the signal from the input side shown in solid blue. (c) Snapshot of the amplitude of integrated pressure at time $t=1.72$~ms.
{(see movie at [Multimedia 1])} }
\label{figapl14khz}
\end{figure}

\begin{figure}[h!]
\includegraphics[width=7.5cm]{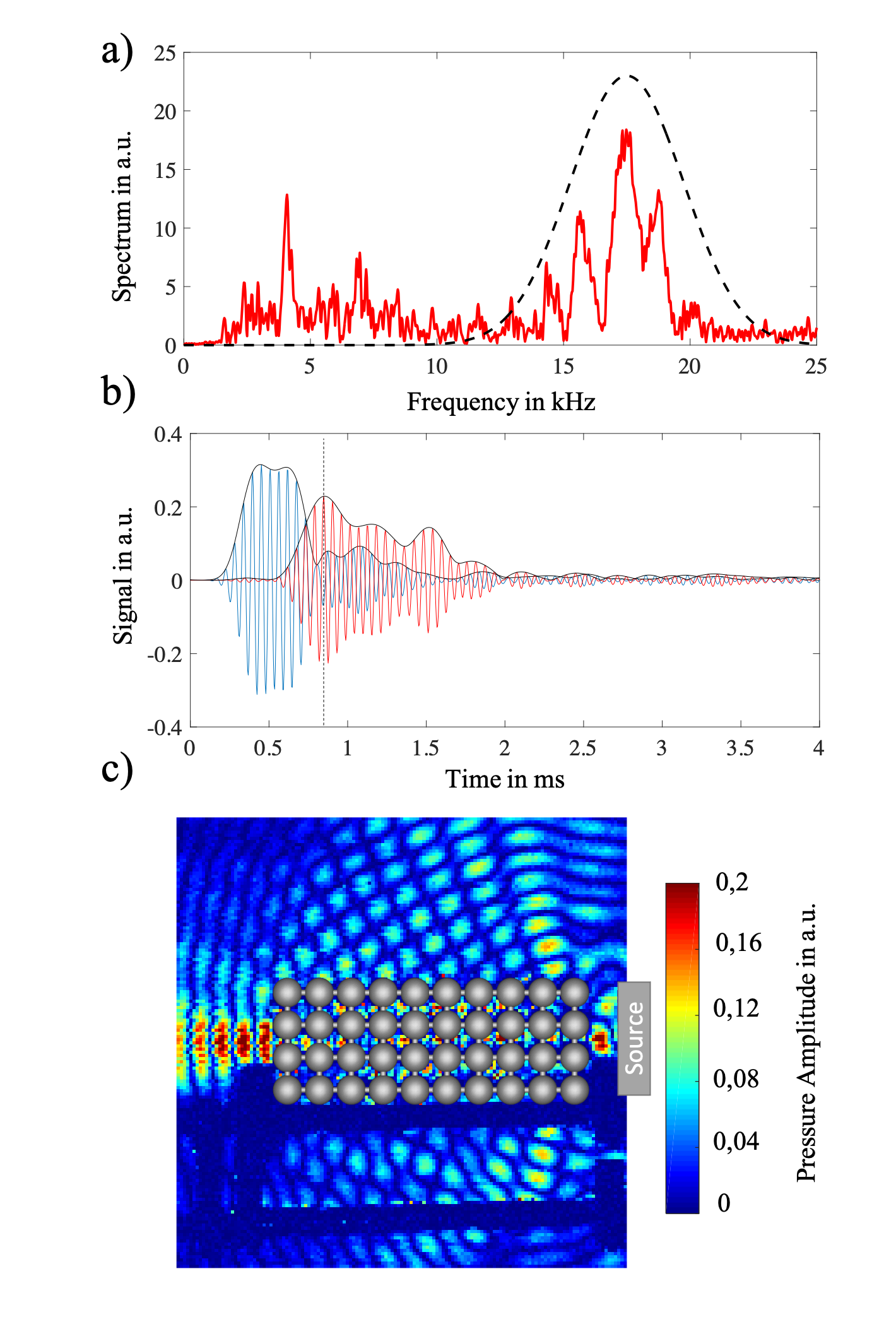}
\caption{Experimental observation at 18~kHz of the endoscopic effect: (a) Spectrum of the transmitted signal for an excitation centered at 18~kHz; the dotted line denotes the Gaussian window used. (b) Transmitted time trace in solid red with the signal from the input side is shown in solid blue. (c) Snapshot of the amplitude of the integrated pressure at time $t=0.88$~ms. {(see movie at [Multimedia 2])}}
\label{figapl18khz}
\end{figure}

To visualize the acoustic propagation in a non-intrusive manner we use a methodology based on the observation of the refractive index change in the air induced by pressure changes    \cite{dubois2019acoustic}. This acousto-optic effect is captured through a compact interferometric sensor designed for this measurement with the  principle first described in~\cite{bertling_OE_2014}. It consists of a commercial laser diode (Thorlabs L1550P5DFB) with single-mode emissions at $\lambda$=1550~nm that points at a reflector so that a part of the emitted light re-enters the laser's cavity producing interferences between the inner optical wave and the back-reflected one. This interferometric approach is known as Self-Mixing Interferometry (SMI)~\cite{donati2011,taimre2015} and it is sensitive to any changes in the time of flight in the external cavity that realizes the laser and the reflector. In the context of acoustic sensing, this change is due to the acousto-optic effect that links the refractive index of air to the pressure among other parameters~\cite{ciddor1996}. With pressure levels encountered in the acoustic domain, the relationship change in refractive index with pressure can be considered as perfectly linear. As expressed in~\cite{bertling_OE_2014} the laser emitted power variations $p(t)$ and the refractive index ones $\delta n(t)$ along the laser propagation axis $z$ are linked as 
\begin{equation}
    p(t)=P_0\cos\left(\frac{4\pi\nu}{c}\int_0^{L_{\rm ext}}\delta n(t,z){\rm d}z + \Phi\right),
\end{equation}
where $P_0$ is the average emitted power, $\nu$ is the laser frequency, $L_{\rm ext}$ is the length of the external cavity and $\Phi$ is a constant phase term.
A part of the light emitted by the laser diode is collected using the monitoring photodiode that is included in the laser package and the photogenerated current is then amplified and converted to a voltage variation that is acquired using a National Instrument acquisition card. This terminal voltage is then an image of the integration of the pressure field along the laser axis in function of time. Images of the acoustic wave propagation come from  the sensor (laser + reflector) which is translated with a fixed step of 1.5~mm on both $x$ and $y$ axes; each position of the sensor represents a pixel of the image shown in Figs.~\ref{figapl14khz}c and~\ref{figapl18khz}c. The experiment is repeated for each laser position in order to represent the integrated pressure field in two-dimensional maps.

%\subsection{Experimental results}
For the temporal results shown in Figs.~\ref{figapl14khz}a,b  and~\ref{figapl18khz}a,b
 the experimental data is temporally downsampled by a fourfold factor leading to a time resolution of 4 $\mu$s. It is then filtered spectrally with a Gaussian window to obtain the acoustic response with chosen central frequency and bandwidth. The Gaussian window presents a full width at half maximum of 5.25 kHz and is centered in 14.5 kHz and 17.7 kHz as shown in Figs.~\ref{figapl14khz}a and~\ref{figapl18khz}a. 

%\begin{figure}[h!]
%\includegraphics[width=7cm]{slideMarc_iso.png}
%\caption{New experimental figure here for comparison with Figures 5 and 6.}
%\label{figapl18khzA}
%\end{figure}

\section{Simulation using a higher-order multi-scale Finite Element Method}
\label{sec:HMsFEM}
\subsection{Method}
We use the higher-order multi-scale Finite element method (HMsFEM \cite{hmsfem_acoustic,hmsfem_elastic}) to rapidly solve the three-dimensional Helmholtz equation for the finite array excited by a source. The underlying methodology is to solve the Helmholtz equation 
in a relatively coarse grid
%(cf. Fig. \ref{fig:mesh_sphere})
and exploit carefully constructed multi-scale basis functions that capture local heterogeneity information and wavenumber information; \cite{hmsfem_acoustic,hmsfem_elastic} compare and contrast HMsFEM with classical  polynomial-based finite element method (FEM) highlighting the advantages in terms of speed and accuracy. A major advantage of HMsFEM is that it significantly reduces the degrees of freedom of the FEM defined on the fine scale mesh and thereby saves both a huge amount of CPU time and memory. The 
construction of the multi-scale basis in a coarse element, $K$,
%(see Fig. \ref{fig:mesh_sphere})
for two types of multi-scale basis associated with the 
boundary and interior interpolation points of the polynomial basis functions
is the key building block required for HMsFEM. The first type is obtained by solving
\begin{equation}\label{fig:pb1}
    \begin{aligned}
        \Delta \phi_i+\Omega^2 \phi_i & =0,\quad \text{in } K,\\
        \phi_i &=\Phi_i, \quad \text{on } \partial K,
    \end{aligned}
\end{equation}
in which $\Phi_i$ (a Dirichlet datum) is the polynomial shape function
corresponding to the $i^\text{th}$ interpolation point that lies on the boundary $\partial K$ of $K$.
%(highlighted in red in Fig.~\ref{fig:mesh_sphere}). 
For the second type of multi-scale basis, one needs to solve
\begin{equation}\label{fig:pb2}
    \begin{aligned}
        \Delta \phi_j+\Omega^2 \phi_j & =\Phi_j,\quad \text{in } K,\\
        \phi_j &=0, \quad \text{on } \partial K,
    \end{aligned}
\end{equation}
where $\Phi_j$ (now a volumetric forcing) is the interior  polynomial shape function associated with the $j^\text{th}$ interpolation point that lies inside of $K$.

The power of the HMsFEM approach is that we replace each standard polynomial basis function in the classical FEM with a multi-scale basis function constructed by solving a Helmholtz equation dependent upon the local problem. Since the basis functions are constructed in a local domain, the computational cost is affordable. Importantly, these multi-scale bases are defined in a very coarse mesh, therefore, the degrees of freedom of the resulting linear system with HMsFEM is greatly reduced compared to a classical FEM. This enables us to solve the three-dimensional sphere array, with close spacing, both accurately and rapidly. In our numerical simulations, the size of each coarse element ($1.5\times 1.5\times 1.5$~cm) is set to be exactly the same as the cubic cell that contains only one polymer sphere and the computational domain consists of $20\times 10\times 20$ coarse elements. Therefore the size of the simulation domain is $30\times 15\times 30$~cm. Here, for simplicity, we use cubic elements to approximate the  spheres and the length of the fine-scale cubic element is $1/60^{\text{th}}$ the length of the coarse-scale element. 
%(see Fig.~\ref{fig:mesh_sphere}). %In future work, we aim to develop adaptive tetrahedron elements in order to resolve more detail.
Perfectly matched layers  \cite{berenger1994perfectly}  with a thickness of one coarse element are implemented to absorb outgoing waves on all sides of the computational domain. The dimension of the final linear system to be solved is merely 119,629 after utilizing the static condensation algorithm~\cite{wilson1974static} and the CPU time for solving this linear system is just under a minute using Matlab. 

%\begin{figure}
%\includegraphics[width=4cm]{coarse_gll.pdf}
%\caption{Interpolation points distribution of the fourth order Legendre polynomials. In this example, the coarse element contains $10\times 10$ fine elements. The black dots represent interior
%points, whereas the black-edged white dots represent boundary points.
%}
%\label{fig:gll}
%\end{figure}

%\begin{figure}
%\includegraphics[trim={2.2cm 1.5cm 0cm %0cm},clip,width=10cm]{meshb.eps}
%\caption{An illustration of the coarse grid mesh (red) and fine grid mesh (black) in the two-dimensional case. $K$ is an example of coarse element in which the multi-scale basis functions are constructed. 
%}
%\label{fig:mesh2d}
%\end{figure}
%\begin{figure}
%\includegraphics[trim={2.2cm 1.5cm 0cm 0cm},clip,width=10cm]{meshc.eps}
%\caption{An illustration of the  coarse element that includes a sphere in our numerical simulations. The fine scale mesh size is $1/60$ of the coarse scale mesh size.  
%}
%\label{fig:mesh_spere}
%\end{figure}

\iffalse
\begin{figure}
\includegraphics[trim={2.2cm 1.5cm 0cm 0cm},clip,width=8cm]{meshc.eps}
\caption{
An illustration of the coarse element that includes a sphere in our numerical simulations. $K$ is an example of a single coarse element in which the multi-scale basis functions are constructed. The fine scale mesh size (black) is $1/60$ of the coarse scale mesh size (red).  
\textcolor{red}{Perhaps this figure can be removed as the text clearly explains the situation ? MD}}
\label{fig:mesh_sphere}
\end{figure}
\fi

 \subsection{Results}
\begin{figure}
%\vspace{2.0cm}
\includegraphics[width=8cm]{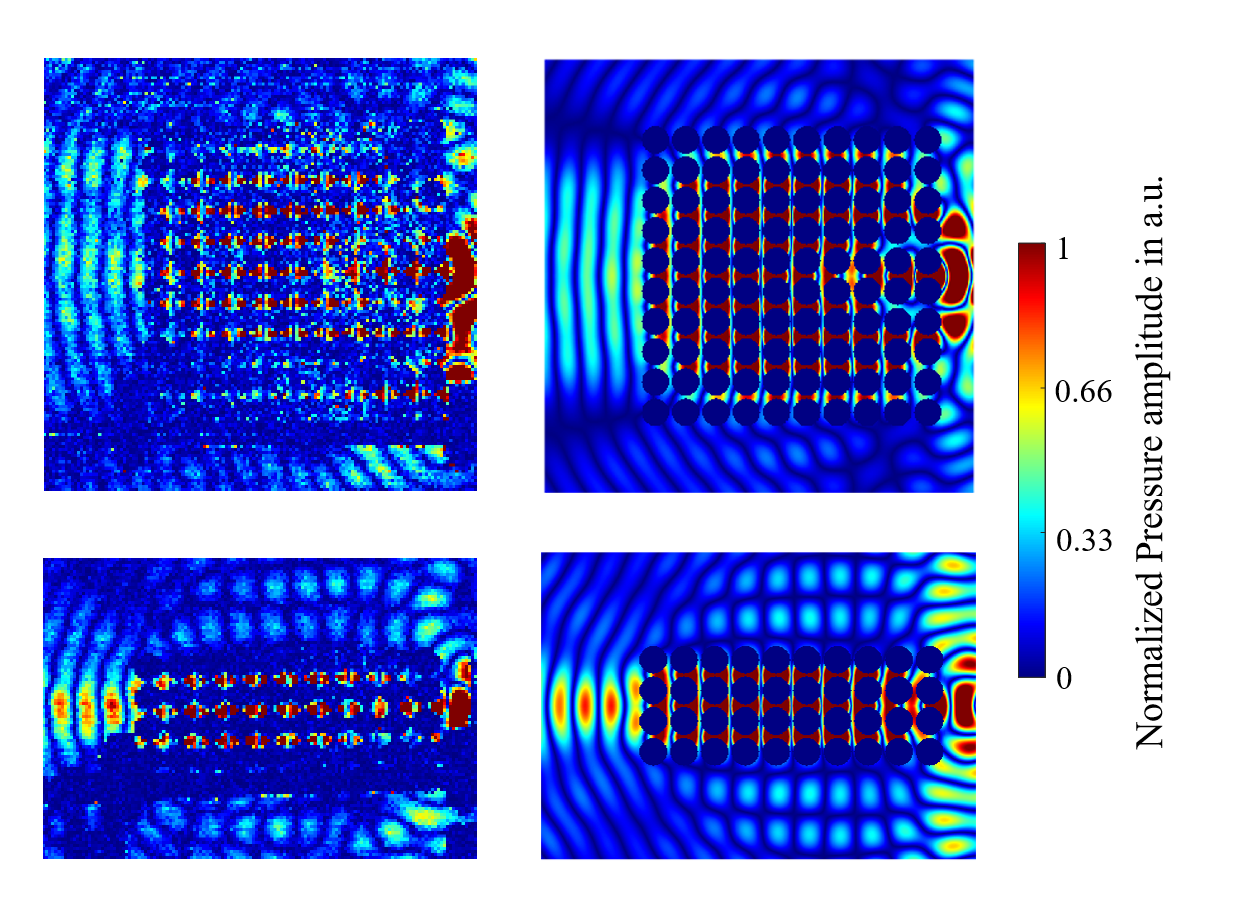}
\caption{Comparison between experiments (left panels) and numerical simulations (right panels) at 14.2~kHz from top (upper panels) and side (lower panels) view.
}
\label{fig:comparison14}
\end{figure}

\begin{figure}
%\vspace{2.0cm}
\includegraphics[width=8cm]{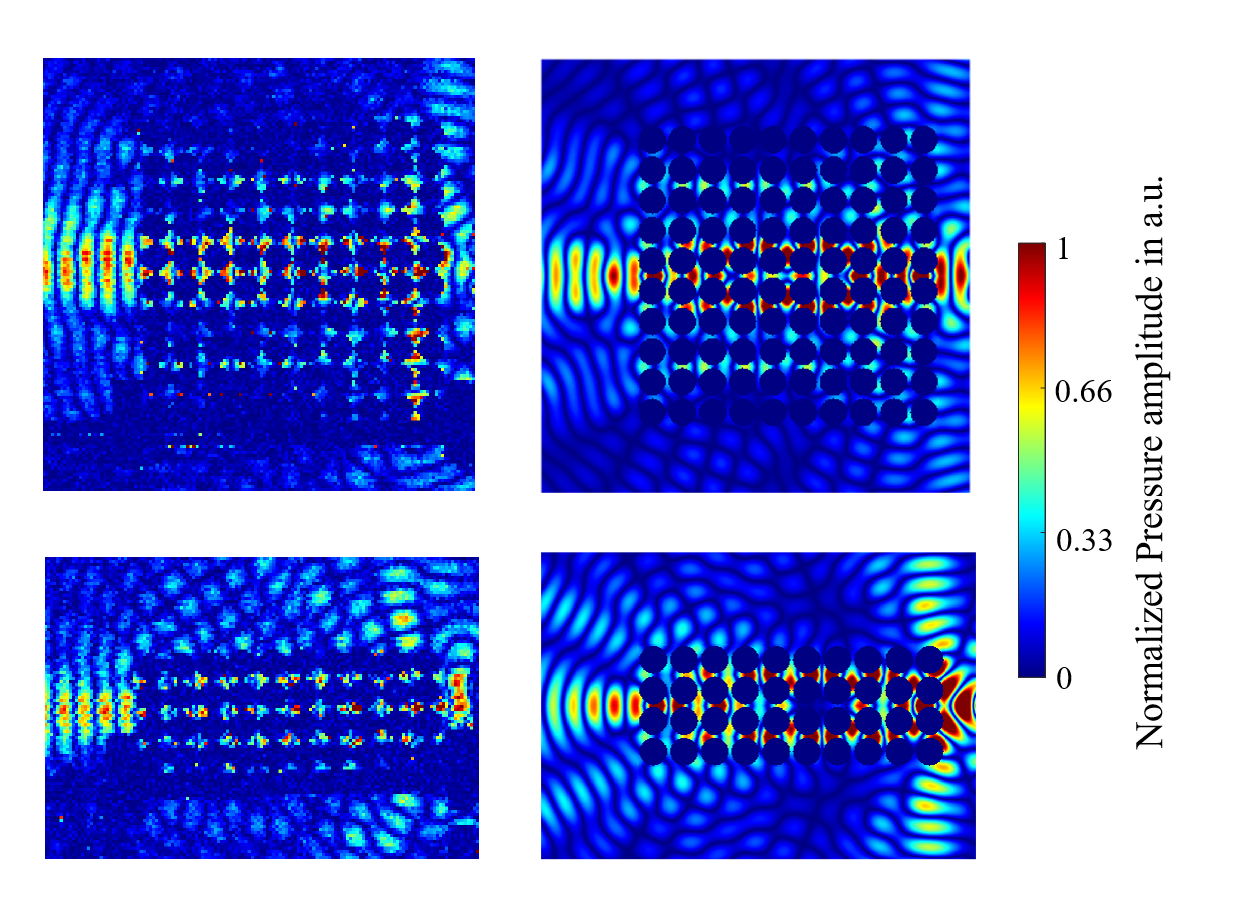}
\caption{Comparison between experiments (left panels) and numerical simulations (right panels) at 18~kHz from top (upper panels) and side (lower panels) view.
}
\label{fig:comparison18}
\end{figure}

The results of both the physical experiments and the HMsFEM numerical simulations for 14.2 and 18~kHz are shown in Figs~\ref{fig:comparison14} and~\ref{fig:comparison18}, respectively; the left columns correspond to the experiments, whereas the right ones show the numerics. The upper rows correspond to the top view of the crystal ($x_1x_2$ plane), the lower row to the side view ($x_2x_3$ plane): there is remarkable agreement between the numerical simulations and the experiments. Both frequencies demonstrate distinctive collimation effects, at 14.2~kHz we see a plane wave exit the crystal that is roughly the width of the crystal, whereas at 18~kHz we have a focused beam exit the crystal.

 %\begin{figure}
 %	\centering
 %	\subfigure[14.2kHz]{
 %		\includegraphics[trim={0.5cm 1.5cm 0.5cm 2cm},clip,width=1.6in]{u142_12}}
 %	\subfigure[14.2kHz]{
 %		\includegraphics[trim={0.5cm 0.5cm 0.5cm 2cm},clip,width=1.6in]{u142_13}} 		
 %	\subfigure[14.2kHz]{
 %		\includegraphics[trim={0.5cm 1.5cm 0.5cm 2cm},clip,width=1.6in]{u142_23}}
 %	\subfigure[14.2kHz]{
%	\includegraphics[trim={1.5cm 2cm 1.5cm 1.0cm},clip,width=1.6in]{u_142_3d}}	
 %	\caption{Endoscopic effect revealed by 3D multi-scale Finite Element computations at $14.2$ kHz.}%%%% Point source results are also available.
 %	\label{figaplFEM14}
 %\end{figure}
 
  %\begin{figure}
 %	\centering
 %	\subfigure[18kHz]{
%	\includegraphics[trim={0.5cm 1.5cm 0.5cm 2cm},clip,width=1.6in]{u18_12}}
 %	\subfigure[18kHz]{
%	\includegraphics[trim={0.5cm 0.5cm 0.5cm 2cm},clip,width=1.6in]{u18_13}}	
 %	\subfigure[18kHz]{
%	\includegraphics[trim={0.5cm 1.5cm 0.5cm 2cm},clip,width=1.6in]{u18_23}}
%	\subfigure[18kHz]{
%		\includegraphics[trim={2cm 2cm 2cm 1.5cm},clip,width=1.6in]{u_18_3d.eps}}
 %	\caption{Endoscopic effect revealed by 3D multi-scale Finite Element computations at $18$ kHz.}%%%% Point source results are also available.
 %	\label{figaplFEM18}
%\end{figure}
 
\section{Periodic acoustic wave problem}
\label{sec:Blochtheory}

We now move to investigate the band spectrum of the sonic crystal since 
 these allow us to look at iso-frequency contours and surfaces and hence gain physical understanding and interpretation of the collimation shown in Figs~\ref{fig:comparison14} and~\ref{fig:comparison18}.
 
%self-guiding properties can be found at the edges of stop bands where the group velocity vanishes \cite{chigrin2003self}.

\begin{figure}
%\vspace{2.0cm}
%\includegraphics[width=9cm]{disp_curves_endoscope.eps}
\includegraphics[width=9cm]{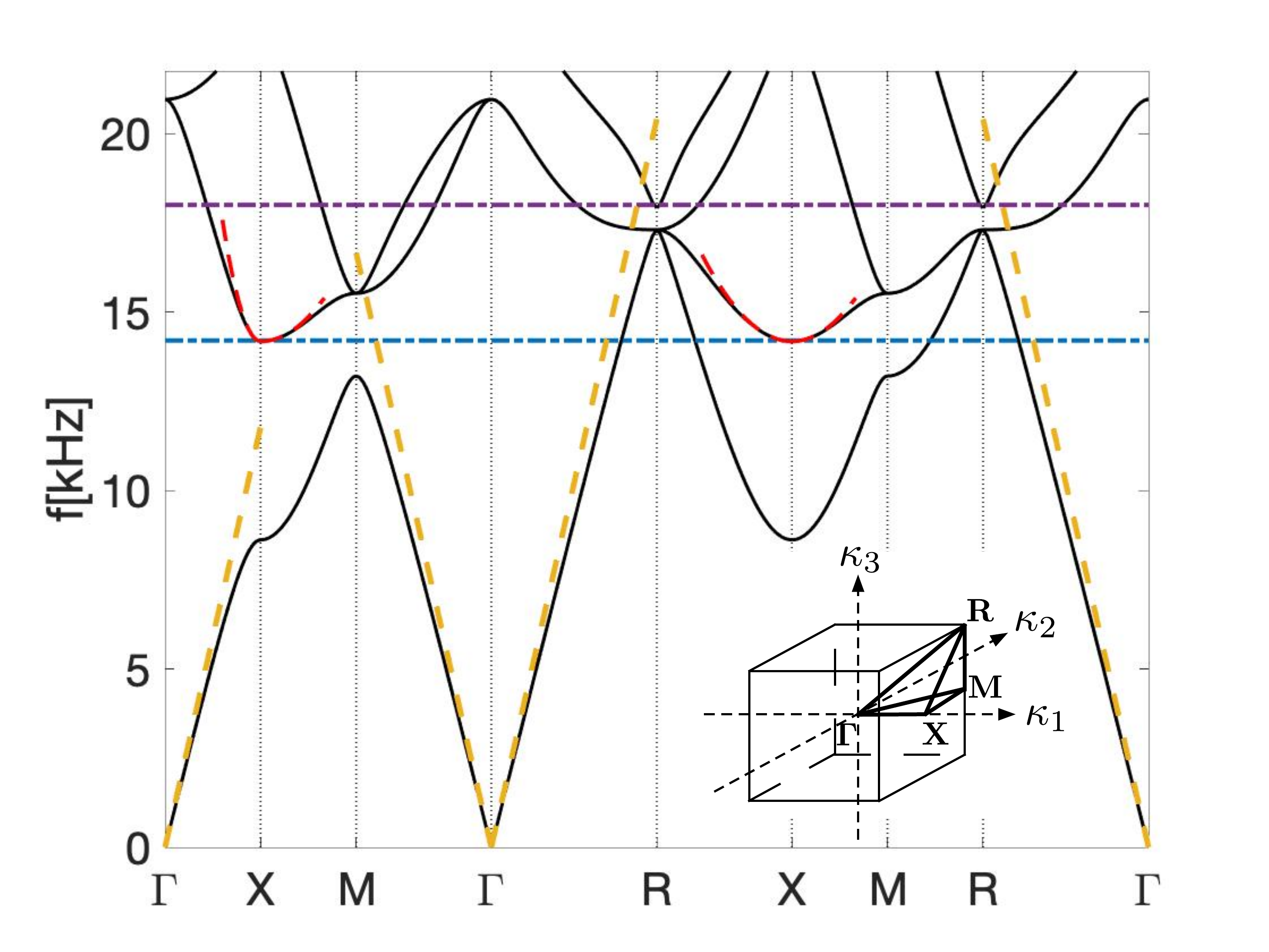}
\vspace{0cm}
\caption{Bloch dispersion curves around the edges of the irreducible Brillouin zone (shown inset) for a cubic array of rigid spheres of diameter $1.38$cm with periodicity $1.5$cm. The global minimum of the second (optical) band at point $X$ corresponds to the frequency of the source in Figs.~\ref{figapl14khz} and~\ref{fig:comparison14} which is plotted as the dash-dotted horizontal blue line at frequency $f=14.2$kHz.
The frequency of the source in Figs.~\ref{figapl18khz} and~\ref{fig:comparison18} is drawn as a dash-dotted horizontal purple line at frequency $f=18$kHz.
The red dashed line show HFH approximations around the $X$ point from Eq.~\eqref{eq:Omega}. The yellow dashed lines represent free-space propagation.
}
\label{fig:apldisp}
\end{figure}

For acoustic waves propagating through an infinite, perfect, triply periodic medium, one can
invoke the Floquet-Bloch theorem \cite{kittel96a,brillouin53a} and 
consider the cubic cell, in non-dimensional coordinates, $-1<x_1,x_2,x_3<1$ with quasi-periodic conditions applied to the
faces. The quasi-periodic Bloch boundary conditions are given as
\beq
  u({\bf x}+{\bf d}_i)=e^{\ri\bm{\kappa}\cdot{\bf d}_i}u({\bf x}),
\quad
\label{Floquet}
\eeq
where ${\bf d}_i$ is the lattice vector along $x_i$, $i=1,2,3$. For instance ${\bf d}_1=(2,0,0)$.
Eq.~\eqref{Floquet} also involves the Bloch wave-vector $\bm{\kappa}=(\kappa_1,\kappa_2,\kappa_3)$
 characterizing the phase-shift as one moves from one cell to the
 next. 
%\textcolor{red}{There is also continuity of $u,~\partial u/\partial x_1,~\partial u/\partial x_2,~\partial u/\partial x_3$ along $x_1=0$, $x_2=0$ and $x_3=0$.} I'm not sure this is needed?
This Bloch problem is solved numerically and the 
dispersion relations that link the frequency $\omega$ and Bloch wavenumber $|\bm{\kappa}|$ are
deduced; as is conventional in solid state physics \cite{brillouin53a}
only a limited range of wavenumbers need be considered in order to detect band-gaps, namely
the wavenumbers along the right-angled tetrahedron $\Gamma XMR$ shown in
the irreducible Brillouin zone (IBZ) in Fig.~\ref{fig:apldisp}; 
 there are some exceptions to this limitation such as operators on graphs \cite{harrison07a}.
 The vertex coordinates are $\Gamma=(0,0,0)$, $X=(\pi/a,0,0)$, $M=(\pi/a,\pi/a,0)$ and $R=(\pi/a,\pi/a,\pi/a)$.

The dispersion curves for a triply periodic array of rigid spheres, shown in Fig.~\ref{fig:apldisp}, illustrate several interesting
features:
a partial stop-band where wave propagation is disallowed along $\Gamma X$ and $XM$ between $13.2$ and $14.2$~kHz, coalescing bands at high-symmetry points $M, X$
and $R$ and regions of flat dispersion curves where the group velocity is zero
and features of slow sound occur~\cite{figotin06a}.
One can see that the group velocity is vanishing at the second band near the high-symmetry point $X$ in Fig.~\ref{fig:apldisp}; the frequency $14.2$~kHz indicated
by the dash-dotted blue line should thus be associated with an interesting wave phenomenon. 
Excitation at, or very close, to the frequency predicted would lead to  oscillations that resemble a standing
wave as it seems to be the case in Fig.~\ref{fig:comparison14}; crucially this standing wave has directionality that can be
identified from the wavenumber description in the Brillouin zone
shown in Fig.~\ref{fig:apldisp}.

%\subsection{High Frequency Homogenization and dynamic anisotropy}
%\label{sec:HFH}
To further confirm the physical mechanism leading to the collimation effect experimentally and numerically observed at $14.2$~kHz, we propose to
replace the sonic crystal by an effective medium in the high frequency regime. The property of this effective medium should depend both on the frequency and the Bloch wavenumber,
as the phenomenon occurs well beyond the quasi-static limit. 

\subsection{Interpretation of collimated beaming at 14.2~kHz with High Frequency Homogenization}\label{sec:HFH}

\begin{figure}
%\vspace{2.0cm}
\includegraphics[width=8cm]{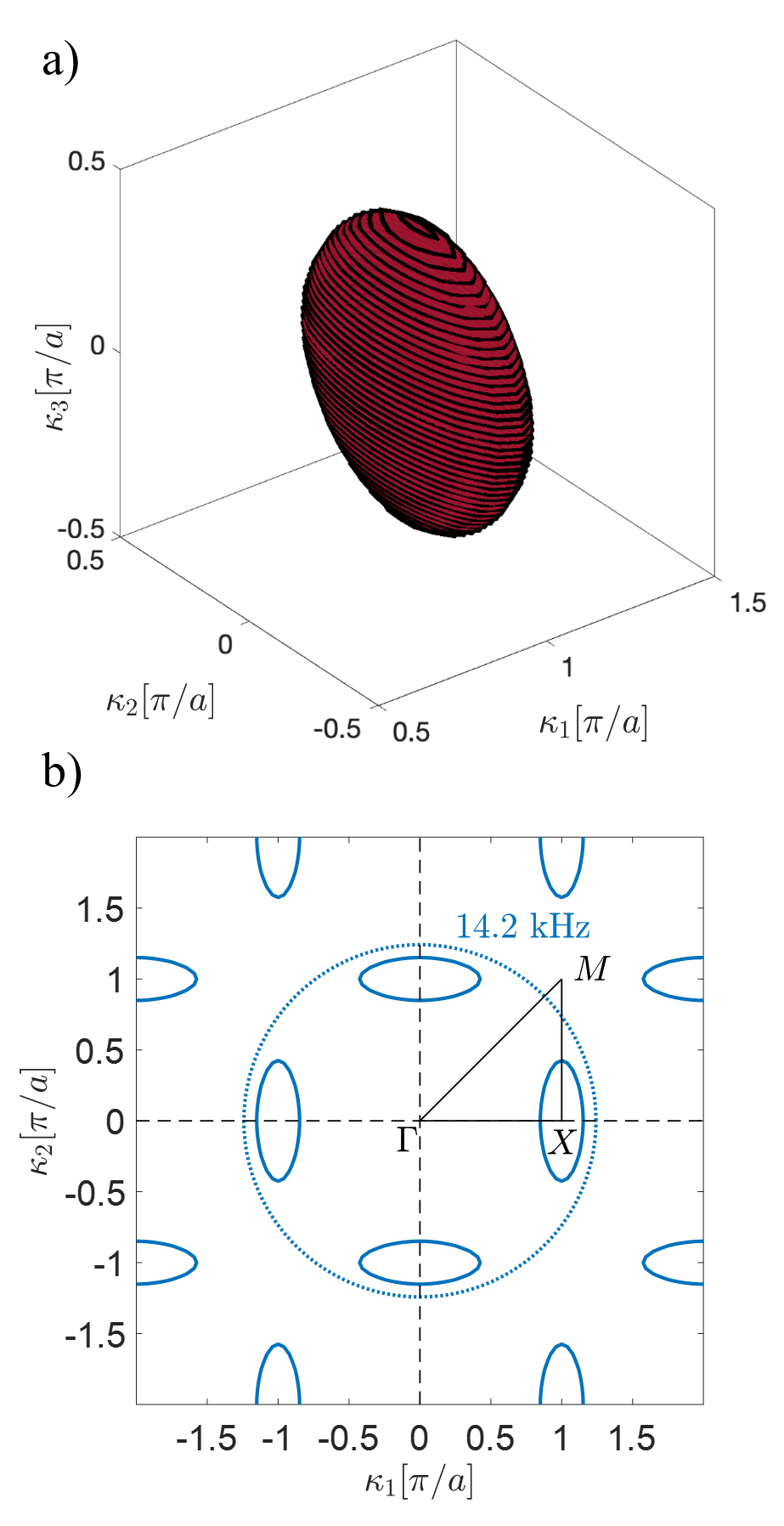}
\caption{Isofrequency contours for the second band at $14.2$~kHz. (a) 3D representation shows an ellipsoid flattened in the $\Gamma-X$ direction. The black lines are obtained with a plane-wave expansion method and superimposed with a red surface derived with HFH. (b) 2D representation of the isofrequency contour in the plane $\kappa_3=0$ with the free-space isofrequency contour at $14.2$~kHz in dashed line (see dotted blue line in Fig.~\ref{fig:iso14}, or yellow on Fig.~\ref{fig:apldisp}).
}
\label{fig:iso14}
\end{figure}

The collimation effect observed at 14.2~kHz is related to strong anisotropy as shown by the equifrequency surface of Fig.~\ref{fig:iso14} which is wrapped around $X$ point with strong ellipticity in $\kappa_1$ and $\kappa_3$ directions. The slight mismatch of frequency to $14.2$~kHz for the infinite periodic crystal is attributed the finite size of the crystal used in the experiment and simulations. 
This leads to a plane wave like behavior in the crystal and results in the wave leaving the crystal with a flat phase profile. This is, in effect, a collimation experiment where the compact source is transformed into a plane wave similar to the Luneburg lens based on index gradient media  \cite{climente2010sound,di2011luneburg,chang2012enhanced,romero2013wave,zhao2016anisotropic}. The collimation effect is related to the excitation of a single Bloch mode of the crystal leading to a plane wave like propagation in the crystal as illustrated in Fig.~\ref{fig:comparison14}. This is further confirmed by two experimental observations: (i) we observe a clear single frequency contribution at the exit of the crystal in Fig.~\ref{figapl14khz}a and (ii) the group velocity is significantly reduced as observed in Fig.~\ref{figapl14khz}b with the pulse delay. The movie of the experiment [Multimedia 1] is instructive in showing the combined effect of the low group velocity and the resultant delay allowing the wavefield to spread across the crystal.
 
In Fig.~\ref{fig:apldisp}, we approximate the dispersion curves computed via Finite Elements (Comsol Multiphysics) with
asymptotic curves originated in the High-Frequency Homogenization theory (HFH) developed in \cite{craster10a}.
The multiple scales used for HFH are a short-scale $\xi_i=x_i/l$ and a
long-scale  
$X_i=x_i/L$ for $i=1,2,3$ where $l$ and $L$ represent respectively the 
characteristic small scale (half length of a cell) and the long scale. 
A small parameter is formed as $\epsilon:=l/L$ and expansions of the frequency
$\Omega^2=\Omega_0^2+\epsilon\Omega_1^2+\ldots $ and the solution $u=u_0+\epsilon
u_1+\ldots$ are substituted in~\eqref{eq:helmholtz}. Here $\Omega_0$ is a
standing wave frequency given at the high-symmetry points $X$, $\Gamma$, $M$ and $R$ of the irreducible
Brillouin zone and
a perturbation scheme can be developed about the band edges. The
leading-order term for $u$ is $u_0(\boldsymbol\xi,{\bf X})=f_0({\bf
  X})U_0(\boldsymbol\xi)$ where the function $f_0$ representing an
envelope of the solution $u_0$ is obtained by an equation only on the
long scale: The theory is detailed in  \cite{craster10a}. Changing
back to the original $x_i$ coordinates the effective medium equation
\cite{craster10a} reads,
\beq
T_{ij}\frac{\partial^2 f_0}{\partial x_i\partial x_j}-\frac{(\Omega^2-\Omega_0^2)}{l^2}f_0=0.
\label{eq:f_0}
\eeq
$T_{ij}$'s are integrated quantities of the leading order and
first order short scale solutions with $T_{ij}=0$ for $i\neq j$ in the present illustrations. As in \cite{craster10a},
assuming Bloch waves, the asymptotic dispersion relation for $\Omega$ reads,
\beq
\Omega\sim\Omega_0-\frac{T_{ij}}{2\Omega_0}\kappa_i\kappa_j,
\label{eq:Omega}
\eeq
where $\kappa_i=K_i-d_i$ and $d_i=0,-\pi/2,\pi/2$ depending on the
band edge in the Brillouin zone about which the asymptotic expansion
originates ; these asymptotics give the red dashed curves in Fig.~\ref{fig:apldisp}. For the case of multiple modes originating from the same point, as is the case for the second and third bands at point $M$ and the first and second bands at point $R$, Eq.~\eqref{eq:Omega} is no longer valid and one obtains two coupled equations for $f_0^{(i)}$.
This homogenization theory is not limited to long-waves
relative to the microstructure, one apparent failing is that the
asymptotics appear to be only valid near the band edge frequencies but
further refinements are possible, using foldings of the Brillouin
zone, that extend the theory to provide complete coverage of the
dispersion curves and provide accuracy at all frequencies, see e.g.~\cite{antonakakis2013high} and further developments in~\cite{guzina2019rational,assier2020high}.

\iffalse 

Since the full dispersion curves live in four dimensions for a three-dimensional crystal, the complexity of the complete band structure is better appreciated by drawing isofrequency surfaces. Actually, we remind as a word of caution that a missing mode is almost certainly present in many periodic structures that have already been analysed and has been missed. Such a warning has been made for a doubly periodic array of rigid cylinders in~\cite{craster2012a} and a triply periodic array of rigid spheres in \cite{dubois2019acoustic}. Extreme dynamic effective anisotropy has also been demonstrated for a doubly periodic pinned elastic plate~\cite{lefebvre17}. Thus, this missing mode is a feature that can be encountered in many wave scenarios. In the present case, isofrequency surfaces at $14.2$ and $18$~kHz are shown on Fig.~\ref{fig:iso14} and~\ref{fig:iso18}, respectively.
%The following ansatz are taken in (\ref{eq:helmholtz}) for the eigenfrequencies  
%\begin{equation}
%\Omega^2=\Omega_0^2+\epsilon \Omega_1^2+\epsilon^2 \Omega_2^2+\ldots
%\label{eq:expansion2D}
%\end{equation}
%and associated eigenfields
%\begin{equation}
%u({\bf X},\bxi)=u_0({\bf X},\xi)+\epsilon u_1({\bf X},\xi)+\epsilon^2 u_2({\bf X},\xi)+\ldots
%\label{eq:expansion2D}
%\end{equation}
%
%Since $u({\bf X},\xi)$ is periodic in $\xi$ so are the $u_i({\bf
%  X},\bxi)$'s.

\fi 

%\subsection{Endoscope effect at 14.2 kHz}
Fig.~\ref{fig:iso14} shows an ellipsoid flattened in the $\Gamma -X$ direction. The black lines are obtained from direct computation with Comsol Multiphysics. This isofrequency surface is superimposed by a red surface obtained from HFH. We point out that this demonstrates the sharpness of HFH approximation not only along the IBZ boundary as in Fig.~\ref{fig:comparison18}, but also within the IBZ. The effective medium at $14.2$ kHz is described by the matrix coefficients $T_{11}=T_{33}=31.4$ and $T_{22}=244.4$, which confirms the large anisotropy along the $x_2$ axis in agreement with the observed guiding effect in that direction with the sonic crystal in Fig.~\ref{fig:comparison14}.
Interestingly, the 2D acoustic magnifying hyperlens of Zhang's group consisting of brass fins in air embedded on a brass substrate~\cite{li2009experimental} depicts elliptical isofrequency contours similar to those in Fig.~\ref{fig:iso14}(b) at a working frequency of 6.6~kHz. When viewed from the $\Gamma$ point, elliptical contours within the dashed circle in Fig.~\ref{fig:iso14}(b) appear locally hyperbolic. One indeed observes a collimated beam on the exit of the hyperlens in~\cite{li2009experimental}, analogous to that on the exit of the sonic crystal in Fig.~\ref{fig:comparison14}. In electromagnetism~\cite{lu2006experimental} demonstrate self-collimation for a source inside a finite crystal in the microwave regime where the isofrequency contours are also hyperbolic and here we see a similar effect, but now for acoustic waves.

\subsection{Interpretation of collimated beaming at 18~kHz with Fermi surfaces and slowness curves}
\begin{figure}
%\vspace{2.0cm}
\includegraphics[width=8cm]{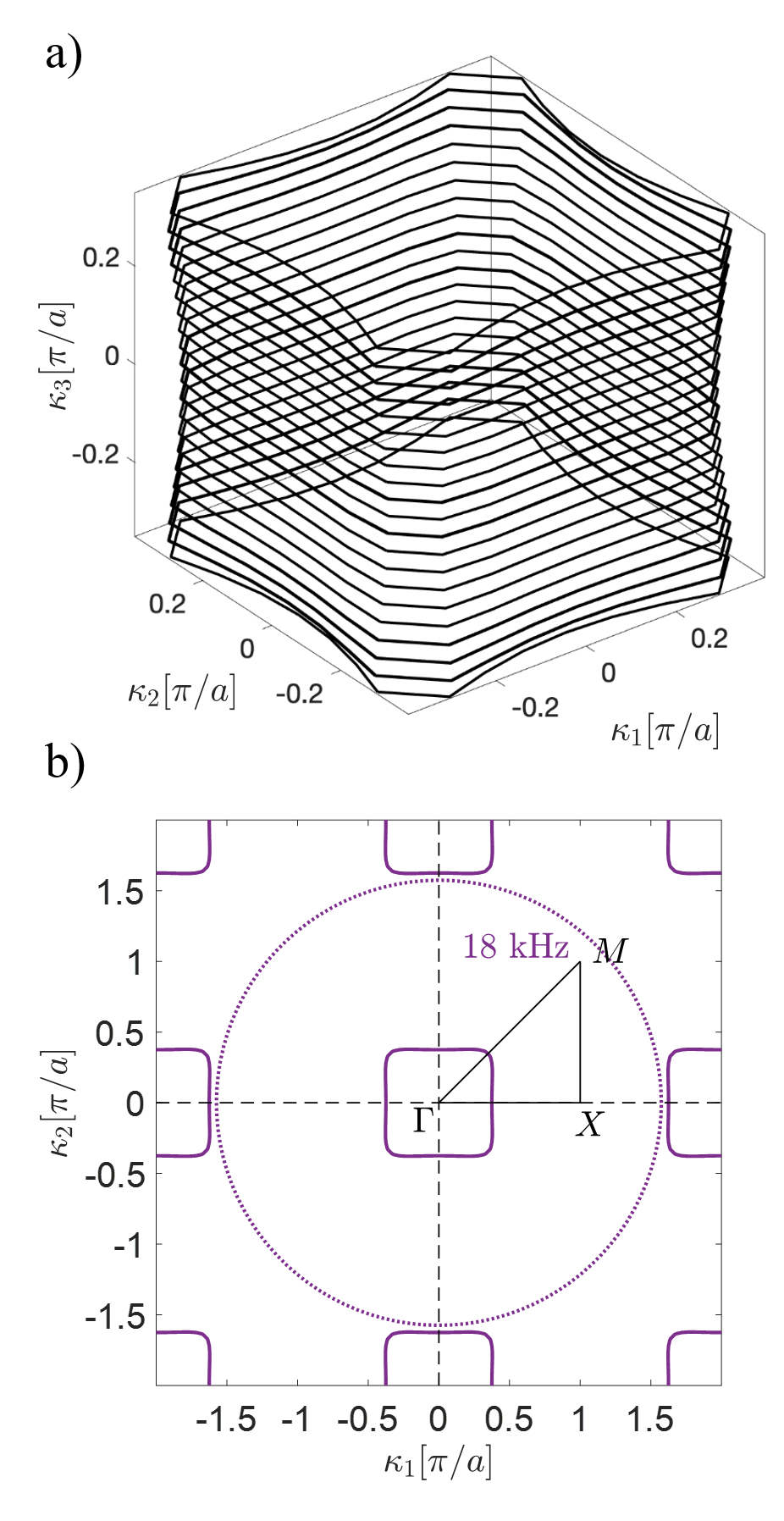}
\caption{The isofrequency contours for the second band at $18$~kHz. (a) 3D representation shows concentric near-squares centred around $\Gamma$ and flat perpendicular to the $\Gamma-X$ direction. (b) 2D representation of the isofrequency contour in the plane $\kappa_3=0$ with the free-space isofrequency contour at $18$~kHz in dashed line (see dotted purple line in Fig.~\ref{fig:iso18}, or yellow on Fig.~\ref{fig:apldisp}).
}
\label{fig:iso18}
\end{figure}

We directly use the isofrequency curves for this case as they give a self-explanatory insight of the physics. Indeed, by looking at the isofrequency surfaces in Fig.~\ref{fig:iso18} we see that they form a slightly deformed cube. The concentric near-squares have almost flat edges in the $\Gamma-X$ direction 
and notably the isofrequency contour in the plane $\kappa_3=0$ surrounds the $\Gamma$ point in Fig.~\ref{fig:iso18}b leading to anti colinear phase and group velocity (see also movie at [Multimedia 2]). In the crystal this is reminiscent of an endoscope or self-guiding effect where the narrow support of the source is conserved across the crystal and then leaves the crystal as a focused beam.

Unlike the 14.2~kHz observation of collimation, at 18~kHz we experimentally observe the excitation of several Bloch modes in the crystal as expected from Fig.~\ref{figapl18khz}a. The field is dominated by the mode giving these cubic equifrequency surfaces that lead to the self-guiding effect in the crystal. We note the similarity of the isofrequency contour at $18$ kHz with that in \cite{chigrin2003self} that leads to a self-guiding effect in 2D photonics whereas we propose a 3D experimental realization of the effect in acoustics. We further refer to \cite{Prather2005SelfcollimationI3} for validation in the microwave regime of self-collimation and to \cite{lu2005three} for a 3D focusing effect via all-angle-negative refraction.

%\begin{figure}
%%\vspace{2.0cm}
%\includegraphics[width=8cm]{contours_at_0.eps}
%\caption{Isofrequency contours of the second band at $14.2$ kHz (blue) and $18$ kHz (purple) in the plane $\kappa_z=0$.}
%\label{fig:contours_0}
%\end{figure}

\section{Concluding remarks}
\label{sec:conclude}

%\begin{figure}[h!]
%\includegraphics[width=8cm]{slideMarc_iso.png}
%\caption{When approaching 14kHz and thanks to the high resolution of the experiments, we clearly see that there is a change in the pressure field distribution in the crystal. It looks like the maximum as shifted by half a period, now presenting a line of 0 pressure amplitude between the scatterers. Marc is wondering if this effect could be attributed to the equifrequency contour that end up wrapped around the edge of the Brillouin zone before the first collimation is observed around14.5kHz. See also movies at https://amubox.univ-amu.fr/s/ZxjcJP5P3YwLxF6}
%\label{figapl18khz2}
%\end{figure}

We have conclusively shown strongly collimated beaming effects through a three-dimensional acoustic phononic crystal consisting of a finite rectangular cuboid array of $4\times 10\times 10$ polymer spheres $1.38$~cm in diameter in air, arranged in a primitive cubic cell with pitch $1.5$~cm for chirps with central frequencies at $14.2$~kHz and $18$~kHz. Our study is making use of three-dimensional higher-order multi-scale finite element simulations that enable us to capture the essence of the wave phenomena for our scattering problem in the time domain. These simulations show excellent agreement with time-domain experiments based on an in-house laser feedback interferometer setup. The prediction of the two frequencies where these effects are achieved requires analysis of band spectra of a three-dimensional periodic structure, that reveals the prominent role played by the Floquet-Bloch waves in the infinite periodic counterpart of the finite rectangular cuboid array of rigid spheres. Visualization of four-dimensional Fermi surfaces in the infinite crystal is somewhat delicate and thus we further analyse effective properties of the periodic structure through the lens of a high-frequency homogenization method. Our analysis shows one needs to be extra-careful when
just considering edges of the three-dimensional Brillouin zone instead of its volume. While Fermi surfaces
(here 4D dispersion surfaces) contain within them all the information required to
completely describe the stop band structure of a 3D periodic structure, dispersion
curves such as in Fig.~\ref{fig:apldisp} can obscure important features such as the remarkable isofrequency contours in Fig.~\ref{fig:iso14}-\ref{fig:iso18}, and most notably at $18$~kHz wherein the phase and group velocity are anti-linear.
This a frequency regime where the dispersion curves in Fig.~\ref{fig:apldisp} are difficult to decipher.
Moreover, it is not obvious to conclude from  Fig.~\ref{fig:apldisp} that there should be highly directive wave propagation within the crystal at $14.2$~kHz, as evidenced by Fig.~\ref{fig:comparison14}. On the other hand, the combination of HFH that reveals strong effective anisotropy of the crystal through inspection of the $T_{ii}$ coefficients deduced from Eq.~\ref{eq:f_0}, and of the isofrequency contours that display a remarkably flat isosurface in Fig.~\ref{fig:iso14}, sheds light on the waveguiding effect through the phononic crystal at $14.2$~kHz. The sonic crystal can be viewed as a waveguide counterpart to a uni-directive metamaterial antenna with a vanishing effective refractive index such as in~\cite{enoch02a}. It is well known that curved wave fronts emanating from a source in free space are transformed into planar ones in a zero index slab~\cite{ziolkowski2004propagation,liberal2017near}. 
In the present case, the combination of low group velocity in the $X_1$ direction and large dynamic anisotropy of the sonic crystal at $14.2$~kHz makes it possible to design an effective highly anisotropic low-index waveguide.
%However, such an ultra-refraction route towards a collimated beaming seems less straightforward for the acoustic setup.
Quite unexpectedly for such a simple phononic crystal consisting of a finite rectangular cuboid array of rigid spheres, it turns out that isofrequency contours can design for practically implementable collimated beams at two seemingly non remarkable frequencies, at least according to the dispersion diagram. It is only through further inspection of isofrequency surfaces and contours, in conjunction with high frequency homogenization, that the true nature of the wave propagation within the crystal is unveiled at these frequencies. Our study thus emphasizes the richness of physical phenomena via anomalous dispersion in 3D phononic crystals. We hope our study will foster experimental efforts in manipulation of sound in miscellaneous phononic crystals whose dynamic effective properties of practical importance for wavefront shaping might have been overlooked.

\section*{Acknowledgments}
K.B. and A.D.R. acknowledge financial support provided by the Australian Research Council (DP210103342). G.L. acknowledges support from Newton International Fellowships Alumni following-on funding awarded by The Royal Society. M.D., R.A. and S.E. acknowledge financial support from the Excellence Initiative of Aix-Marseille University - A*MIDEX, a french “Investissements d'Avenir” programme under the Multiwave chair of Medical Imaging.
\medskip

MULTIMEDIA link (https://amubox.univ-amu.fr/s/MWfxJG5ZTzaNsnP)

%The lost boy can be after all a trivial comedy
%for serious people. If one would like to avoid dealing with 4D dispersion surfaces, such as in 3D mechanical metamaterial \cite{buckmann14}, then our unfolding IBZ method seems to be a good alternative, and besides it is interesting to note it is mathematically justified by earlier work on asymptotic analysis of Bloch wave spectra \cite{allaire98}.

%\textcolor{red}{from Marc : What I'm surprised with is that we also intercept some contours near the R point which is related to the direction along the shortest axis of the lens. I dont know for example if we cant really see the wave propagating on this axis because it is much shorter than the other. I dont know if we checked that with a simulation of a 10 x 10 x 10 sphere crystal instead of 10 x 10 x 4 as it is today ?I dont think we need to add such information to the article but it could be useful to be ready to answer anyway if a reviewer picks on this.}

%\bibliographystyle{apsrmp}
\bibliography{references}

%\section{Supplemental material (potpourri)}
%figure * goes across both columns
%\begin{figure}[h!]
%\includegraphics[width=9cm]{Experimental_results_14khz_Freespace_V2.png}
%\caption{Experimental observation of the endoscopic effect at 14.2 kHz. Top: Snapshot of the pressure amplitude for free space (left) and through the sphere lattice (right). Bottom: Time traces of both cases, the blue line represents the input signal outgoing from the source. The red trace corresponds to the pressure signal on the image side of the structure. Vertical dotted lines denote snapshots times for both cases (1ms and 1.63ms respectively). The position of the source and a schematic of the sphere lattice is superimposed to the experimental measurement.}
%\label{figapl1}
%\end{figure}

%\begin{figure}
%\vspace{2.0cm}
%\includegraphics[width=8cm]{figyounes.pdf}
%\caption{{\bf for supplemental} Emission of an acoustic at normalized frequency $\Omega=2.24$, placed in the center of a $10\times 10\times 10$ array of rigid spheres. {\bf We need this at the global minimum of the second band and also for the scattering problem (this will show the wave trajectories in the sonic crystal leading to the endoscope effect)}.
%}
%\label{figapl4}
%\end{figure}

\end{document}